\documentclass[%
 reprint,
superscriptaddress,
 amsmath,amssymb,
 aps,
longbibliography,
prb,
floatfix,
]{revtex4-2}

\usepackage{graphicx}
\usepackage{dcolumn}
\usepackage{bm, bbm}
\usepackage{xcolor}
\usepackage{float}

\begin{document}


\title{Stochastic Metholodgy Shows Molecular Interactions Protect 2D Polaritons}

\author{Nadine C. Bradbury}
\affiliation{Department of Chemistry and Biochemistry, University of California, Los Angeles, CA, 90095, USA}
\author{Raphael F. Ribeiro}
\affiliation{Department of Chemistry and Cherry Emerson Center for Scientific Computation, Emory University, Atlanta, 30322, GA, USA}
\author{Justin R. Caram}
\email{jcaram@chem.ucla.edu}
\affiliation{Department of Chemistry and Biochemistry, University of California, Los Angeles, CA, 90095, USA}
\author{Daniel Neuhauser} 
\email{dxn@ucla.edu}
\affiliation{Department of Chemistry and Biochemistry, University of California, Los Angeles, CA, 90095, USA}


\date{\today}

\begin{abstract}
We introduce stochastic techniques that enable the simulations of polaritons resulting from placing giant 2D molecular aggregate crystals with $10^8$ interacting excitonic dyes in realistic multi-mode cavities.  We show that  the intermolecular coupling protects the formation of polariton states in the face of strong molecular disorder due to persistent delocalization of the dark molecular states. This demonstrates the nontrivial role of internal aggregate Hamiltonian in polariton properties, and the new computational method opens horizons for stochastic simulations of related systems.

\end{abstract}

\maketitle



 Experiments have shown that photophysical and even chemical properties can be modulated by optical cavities, leading to promising potential applications across chemical and materials science domains.~\cite{Coles2014,Reitz2018, DelPo2021, Zhong2016,Zhong2017,MartnezMartnez2018}  Thus far, many different types of substrates have been demonstrated with strong electronic coupling to cavities, including semiconductor crystals, molecular aggregates, and organic polymers.  Despite this breadth, treatments of experimental data typically rely on the simple Tavis-Cummings Hamiltonian,\cite{Tavis1968} which neglects direct intermolecular interactions between emitters. Any complete description of light-matter interactions should account for the often complex DOS availed by extended matter.

Furthermore, even when multiple molecules are considered, most theoretical studies of molecular polaritons only represent the electromagnetic field with a single boson mode.  For system sizes up to a few dozen molecules, high level theoretical methods accurately reproduce simple optical observables.~\cite{Tichauer2022,CamposGonzalezAngulo2022,Chuang2022front}. However, with the inclusion of long-range coupling, `giant' systems are needed to predict accurate delocalization and transport properties.

Intermolecular interactions and multiple photonic modes are especially important for  molecular aggregates and related biological light harvesting systems-- all of which have strongly internal-structure-dependent collective superradiant properties. This is especially true in J-aggregates, one of the first studied systems that can form molecular optical polaritons.\cite{Kasha1963,Lidzey1998} The energy transfer properties in J-aggregates relate to internal geometry and the corresponding electronic band structure.~\cite{Caram2016} Interestingly, tight-binding models suggest strong light-matter coupling can be employed to manipulate spectral and transport properties of dark exciton states (i.e. states with low photonic content).~\cite{Botzung2020, Scholes2020,Ribeiro2022,Allard2022,Engelhardt2022,Zhou2023,Aroeira2023}

The importance of using a full multi-mode cavity is highlighted in recent works.~\cite{Agranovich2003,Michetti2005,Agranovich2007,Ribeiro2022,Mandal2023,Engelhardt2023, Aroeira2023} For example, it was shown in red detuned devices the anticrossing between optical and exciton modes is shifted to higher wavevectors, protecting a greater fraction of lower polariton states from localization induced by static molecular disorder.~\cite{Michetti2005, Litinskaya2006,  Ribeiro2022, Suyabatmaz2023} Thus, a many-mode cavity representation is essential for a rigorous investigation of disorder-resistant transport in polaritonic materials.

Here we apply an extremely efficient linearly-scaling stochastic approach to study polaritons in large micron-sized multi-mode cavities containing two-dimensional (2D) molecular aggregates/crystals with tens of millions of molecules. Stochastic trace techniques\cite{Bradbury2020} are used to visualize the density of states (DOS), participation ratio, and angle dependent transmission. Our main finding is that the aggregate structure drastically affects the disorder-dependent properties of the resulting cavity-induced polaritons and weakly coupled states, including lineshapes and delocalization measures. These results reinforce that inaccurate characterization of the intermolecular interactions will yield poor results in describing photophysical and transport properties of molecular agreggates in the strong coupling regime.

\begin{figure}
    \centering
    \includegraphics{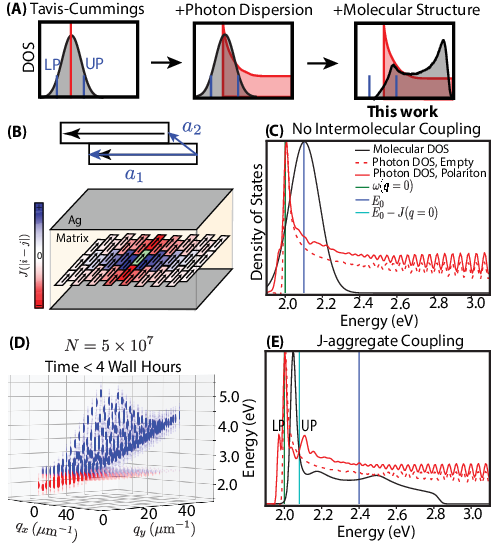}
    \caption{(A) Cartoon diagrams displaying the increase in Hamiltonian complexity in the DOS in this work. Black is used for the molecular DOS, red for photon DOS, and blue to label the $q=0$ polariton states. (B) System geometry diagram shows the Coulomb coupling function, $J$, displayed as the coupling to the center green dipole. (C) and (E) DOS diagrams for a J-aggregate and a crystal without Coulombic intermolecular coupling. A macroscopic number of molecules is used, $N_x = 84375, \, N_y = 25$, with a total of 65-75 photons mode  along the long crystal axis below a cutoff of 5.5 eV. The Rabi splitting is around $\sim 0.07$ eV, the energetic disorder (Gaussian) standard deviation is of $0.1$ eV and the Chebyshev resolution is of $0.01$ eV. (D) 3D plot of the angle resolved absorption spectrum of a molecular aggregate measuring 11 x 13 $\mathrm{\mu m}$, including 196 photon wavevectors; modes with energy less than $E_0-J(q=0)$ are colored in red while those above are in blue.}
    \label{fig:agg_cav}
\end{figure}

To show the importance of the intermolecular coupling and its effects on the observables of the traditional Tavis-Cummings Hamiltonian, we employ the  J-aggregate transition dipolar coupling as a minimal model that both shows this effect and has direct experimental parallels. The band-like delocalized density of states in J-aggregates leads to fundamentally different light-matter density of states in the strong coupling regime as shown in the cartoon of Fig. \ref{fig:agg_cav} A. In the SI, we analyze another case, I-aggregation \cite{SeeSI,Deshmukh2022}, where there are dark molecular exciton states lower in energy than the collective aggregate peak at $k=0$, leading again to distinct exciton delocalization properties relative to J-aggregates. The methods presented here are also applicable to any material where the intermolecular interactions are translationally invariant, such as most semi-empirical semiconductor Hamiltonians. 

Our starting point is a dielectric cavity of thickness $L_C$, encompassed by two ideal mirrors.
We only consider the lowest band of photon modes with $q_z = \pi/L_c$, and the $s$ and $p$ polarizations have the same dispersion, $\omega(q_\parallel) = \frac{\hbar c }{\sqrt{\varepsilon_c}}\sqrt{q^2_\parallel + \frac{\pi^2}{L_C^2}}$,
where the zero-wavevector energy, $\omega_0 = \omega(q_\parallel=0)=\frac{\hbar c }{L_C \sqrt{\varepsilon_0}}$, is almost matched to $E_0$, the transition energy of the molecular (dye) exciton, with a detuning $\Delta \equiv \omega_0 - E_0$. The empty cavity Hamiltonian is  ($\hbar=1$)
\begin{equation}\label{eq:Hcav}
    H_C = \sum_{q}\sum_{\lambda=s,p}  \omega(q) a^\dagger_\lambda(q)a_\lambda(q).
\end{equation}

A dielectric slab, either an ordered molecular aggregate or a crystal, with small thickness relative to $L_C$ (Fig. \ref{fig:agg_cav}), is then placed inside the cavity, along its center plane to enhance the light-matter coupling  \cite{Kasha1965,Hestand2018}.  
The dyes are placed on a  2D lattice, with crystal vectors defined as  $a_1 = (0,l)$, $a_2 = (w,s)$, and the lengths of the molecular aggregate are $L_x = wN_x$,  $L_y = lN_y$; the cavity volume is $V_C=L_xL_yL_C$.
On each 2D site $j$ a dye is placed, with a transition dipole $\mu_j$ and an excitation energy $E_j$, shown in Fig. \ref{fig:agg_cav} B.  The dipoles are presumed planar, all pointing in the same direction, here taken to be the $y$ axis, so $\mu_j=\mu_0\equiv\hat{y}$. We assume that the photons and molecule systems share the same Brillouin zone, and share periodic boundary conditions in the plane of the molecule.

We assume energetic site-disorder, $E_j = E_0 + \delta_j$, where $\delta_j$ is  an uncorrelated Gaussian noise with variance $\sigma^2$. Only energetic disorder is considered, rather than positional or orientational, as previous works show that energetic disorder is dominant in molecular aggregates and polaritons.~\cite{Aroeira2023, Ribeiro2022, Doria2018, Litinskaya2006} The fixed-direction dyes interact via a transitionally invariant dipole-dipole interaction, labeled $J_{i-j}$, which to fit experiment is based on finite closely-spaced point-charge interactions.  \cite{SeeSI, Deshmukh2019,Deshmukh2022}

The molecular Hamiltonian is then
\begin{equation}\label{eq:Hmol}
    H_M = \sum_j E_j b^\dagger_j b_j + \sum_{ij} J_{ij}b^\dagger_i b_j.
\end{equation}
With the rotating wave approximation and  the 
Coulomb gauge, the molecule-photon interaction is \cite{Litinskaya2006}
\begin{equation}\label{eq:Hlm}
    H_{MC} = \sum_{j,q}\sum_{\lambda=s,p} [g_{j\lambda}(q)a^\dagger_\lambda(q)b_j + g^*_{j\lambda}(q)b^\dagger_j a_\lambda(q)],
\end{equation}
with a coupling strength
\begin{equation}\label{eq:giq}
    g_{j\lambda}(q) = i \Omega_R \frac{E_j}{E_0}\sqrt{\frac{\omega_0}{N\omega(q)}}p_{\lambda} e^{i r_j\cdot q},
\end{equation}
where the projections along and perpendicular to the field mode are $p_{s} = (\hat \mu_j \cdot \hat{n}_q)$, 
and $p_{p} = (\hat \mu_j \cdot \hat q)$, 
and $\hat n_q = [\hat q \times \hat z]$. The projections are $j$-independent as here all dipoles are parallel. The Rabi splitting strength is 
$\Omega_R = \mu_0\sqrt{\frac{\pi E_0^2N}{\varepsilon_0\omega_0 V_C}}$.\cite{Litinskaya2006} 

\begin{figure*}
    \centering
    \includegraphics[width=\textwidth]{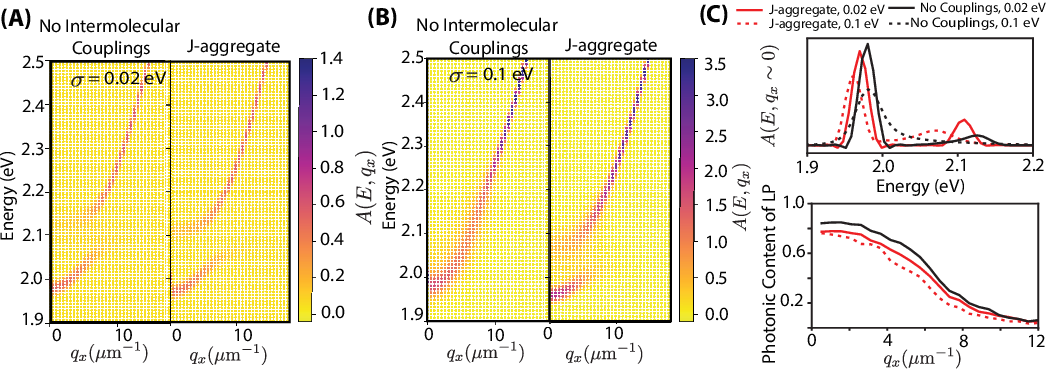}
    \caption{{(A) Angle resolved absorption spectrum of a J-aggregate, and an equivalent (with the same number of monomers) material  with no intermolecular interactions in the weak disorder regime $\sigma/\Omega_R \ll 1$ and (B) strong disorder regime $\sigma/\Omega_R \geq 1$. The photonic content of the LP in (C) was obtained by piecewise integration of the angle resolved photon (transmission spectra) and molecular density of states. Here, the same quasi-1D ribbon was used  as in Figure \ref{fig:agg_cav}, but $N_\zeta =120$ stochastic samples were sufficient to resolve the presented spectra} }
    \label{fig:beta}
\end{figure*}

The full Hamiltonian is then the sum of the cavity, molecular and coupling terms, $H=H_C+H_M+H_{MC}$. Without molecular disorder ($E_j = E_0$) and dye-dye interaction $(J_{ij}=0)$, one obtains the analytically solvable  multi-mode Tavis-Cummings Hamiltonian.~\cite{Agranovich2003} Similarly, in the absence of disorder, we can also exactly resolve the effects of the aggregate internal structure  due to the translational invariance of the Coulomb interaction, which implies  the in-plane wave vector $q$ is a good quantum number for both molecular and cavity subspaces.
In the latter exactly-solvable scenario, using a Fourier-transformed exciton basis,
 $b^\dagger(q) = \sum_j b^\dagger_j e^{iq\cdot r_j}/\sqrt{N}$,
 the Hamiltonian separates into a sum over mode-specific terms, 
\begin{equation}\label{eq:Hq}
\begin{split}
  H_{TOT} = & \sum_q \Big[ E'(q) b^\dagger(q) b(q) +  \sum_\lambda \big( \omega(q)a^\dagger_\lambda(q) a_\lambda(q) \\
    +& g_\lambda(q)a^\dagger_\lambda(q)b(q) 
   + g^*_\lambda(q)a_\lambda(q)b^\dagger(q) \big) \Big] , 
\end{split}
\end{equation}
where the modified exciton energies are $E'(q) = E +J(q)$, with $J(q) = \sum_j J(j) e^{i q\cdot r_j},$ $E(q)=E_0$ is the (constant) exciton energy, while the delocalized exciton-photon coupling term is
$
    g_\lambda(q)  = 2i \Omega_R p_\lambda \sqrt{\frac{\omega_0}{\omega(q)}}.
$
The mode specific exciton-photon Hamiltonian is trivially diagonalized, yielding  polariton states $\xi(q)= \beta(q)b(q) + \sum_\lambda \alpha_\lambda(q) a_\lambda(q) $,  and a simple modification to the
 usual expressions for the upper and lower polariton (UP/LP) energies
\begin{equation}\label{eq:UP/LP}
    E_{UP/LP}(q) = \frac{\omega(q) +E'(q)}{2} \pm \sqrt{\Omega_R^2 + \frac{(\omega(q)-E'(q))^2}{4}}.
\end{equation}

The obtained spectra is exactly the same as that given by the multimode Tavis-Cummings, except that here the momentum-specific interaction replaces the usual non-interacting $E(q)$ molecular energies. For strong interactions as in molecular crystals, with $J(q)$ that may reach up to 0.3eV or more, the Hamiltonian spectrum is substantially modified due to the intermolecular couplings. Note also that as usual, for each polariton the wavefunction amplitudes satisfy:
\begin{equation}\label{eq:alpha_sol}
    |\alpha_\lambda(q)|^2= p_\lambda^2 \frac{(E-E'(q))^2}{(E-E'(q))^2+\Omega_R^2}
\end{equation}
where  $E\equiv E_{UP/LP}$ , while the photon amplitude $\beta(q)$ is determined from the polariton normalization, $|\beta(q)^2|+ \sum_\lambda |\alpha_\lambda(q)^2|=1$. 

We now turn to the nontrivial case of both strong molecular energetic-disorder and intermolecular coupling, $J\sim\sigma\sim\Omega_R$. To model this realistically, we must return to the complete light-matter Hamiltonian. 
The key to efficient very large-scale calculations is the use of stochastic methods, which require that the action of  the Hamiltonian on a given vector be efficient. As the bilinear Hamiltonian conserves the number of polaritons, a single-polariton wavefunction $\psi$ will be a direct sum of a molecular and cavity (photonic) parts, $\psi=\psi^M \oplus \psi^C$, so computationally it is a vector of length $N+2C$, where $N$ and $C$ are the numbers of dye molecules and included cavity modes (the factor of  2 is due to the $s$ and $p$ photon polarizations).  Since the vast majority of photon wavevectors are not in resonance with the excitonic system, an energy cutoff is imposed so  $C\ll N$.  The photon index is denoted by  $\ell=(\ell_x,\ell_y)$, associated with a photon mode $q(\ell)$.

When applying the Hamiltonian on such a function, $H|\psi\rangle $,  the $H_M$ action involves an efficient convolution of the dye-dye interaction,  $\sum_{j\le N} J_{\ell j} \psi^M_j$.~\cite{Bradbury2020} For the cavity-molecule part $H_{MC}$, one can use a similar transform, but it is even faster to apply consecutively fractional 1D Fourier transforms. 
Define $\mathcal{F}_x(\ell_x, j_x) = e^{2\pi i \, j_x\, \ell_x /N_x},$
for elements $ j_x =1 \cdots N_x, \,\ell_x =1 \cdots C_x$ with $\mathcal{F}_y$ analogous.  Then,  for example,
\begin{equation}
\langle \ell \lambda |  H_{MC} | \psi^M\rangle = 2i \Omega_R p_\lambda \sqrt{\frac{\omega_0}{\omega \left(q(\ell) \right) N}}  \mathcal{F}_y \left[ \mathcal{F}_x \left[\frac{E_j}{E_0}\psi^M(j)\right] \right].
\end{equation}
The scaling of this step is $\mathrm{O}(N\sqrt{C})$, and since practical cavities involve at most a few thousand relevant-energy photons, the action of $H_{MC}$ is very efficient.

\begin{figure*}
    \centering
    \includegraphics{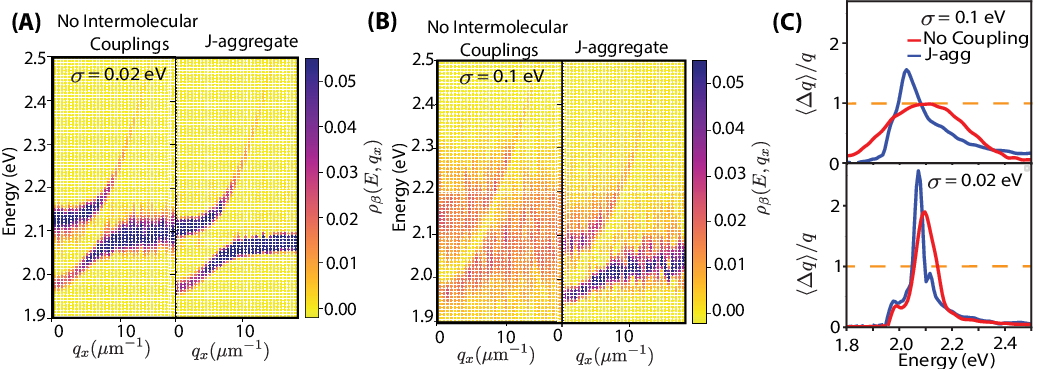}
    \caption{(A) Angle resolved molecular density of states for a J-aggregate and an equivalent system with no intermolecular interactions in the weak and (B) strong disorder regimes. (C) Relative wavevector uncertainty $\langle \Delta q\rangle/q$ of the molecular wavefunctions derived from $\rho_\beta(E,q)$ (Eq. 11) in the case of strong disorder at 0.1 eV (top) and weak disorder 0.02 eV (bottom).}
    \label{fig:ipr_3D}
\end{figure*}

Given the efficient representation of the action of the Hamiltonian, we  use a stochastic resolution of the density of states operator, a common technique in condensed matter systems\cite{Wang1994-JCP, Weisse2006, Bradbury2020}\begin{equation}\label{eq:dos}
    \rho(E) = \mathrm{Tr} [ \delta(H-E)] = \left\{ \langle \zeta |\delta (H-E)|\zeta\rangle \right\} _{\zeta},
\end{equation}
where $\zeta $ is here a vector of length $N+2C$  with random $e^{i\theta}$ elements at each site, and the curly-brackets indicate a statistical average over the random $\zeta$ elements, and simultaneously over the site disorder. The error in the stochastic trace scales as $\mathrm{O}(1/\sqrt{NN_\zeta})$,\cite{Weisse2006} and is thus negligible, for sufficiently large crystals,  even for very few  (here $N_\zeta \approx 10-200$) random samples.  For the action of the DOS operator on a vector, $\delta(H-E) |\zeta\rangle $, the efficient Chebyshev approach is used, \cite{kosloff1988}
with a number of Hamiltonian-vector operations, determined by the desired energetic resolution relative to the spectral width, that is typically less than 2000.~\cite{Bradbury2020}

The overall scaling of the method is then limited by the operations needed to incorporate the intermolecular interactions. In our case the application of the dipolar coupling via convolution is effectively linear in time. In Figure \ref{fig:agg_cav} (D) we show a large calculation possible with this algorithm, requiring only modest computational time that can be parallelized on a standard 128-core AMD Milan cluster. Given the size of the total basis of $10^7$ elements, memory costs associated with wavefunction storage quickly become the limiting computational factor for this method.

To examine the local properties of the molecular and photonic subsystems, we similarly stochastically compute the projected matter and light local DOS, $\rho_{M}(E) \equiv \mathrm{Tr}\left( P_M \delta(H-E)\right)$ and $\rho_C(E) \equiv \mathrm{Tr} \left(P_C \delta(H-E)\right)$,  respectively, introducing the projection operators onto the molecular and cavity(photon) spaces, $P_M$ and $P_C$ respectively. As there are so many more molecules than photon mode involved in strong light-matter coupling, their local density of states are shown separately in Figure \ref{fig:agg_cav}.

The angle resolved photonic density of states (directly proportional to the measurable microcavity angle resolved transmission spectrum), is similarly defined as $A(E, q) \equiv \sum_\lambda \langle q,\lambda|\delta(H-E)|q,\lambda \rangle$, where
$|q,\lambda\rangle \equiv a^\dagger_\lambda(q)|0\rangle $. To efficiently sample it,  we introduce a stochastic resolution of the identity, $\mathbbm{1} = \left\{ |\zeta\rangle\langle\zeta|\right\}_{\zeta}$, which when plugged in yields
\begin{equation}\label{eq:Awk}
    A(E, q) =  \sum_\lambda\left\{\langle q_\lambda |\delta (H-E)|\zeta\rangle \langle \zeta|q_\lambda \rangle\right\}_{\zeta},
\end{equation}
so it is evaluated in the same stochastic process as the total DOS (Eq. 9), as both use the $\delta(H-E)|\zeta\rangle$ vector.

Without static disorder, the angle resolved  transmission is simply proportional to $A^0(E,q) = \sum_{\eta = LP, UP}  \sum_\lambda |\alpha_\lambda(q)|^2 \delta [E-E_\eta(q)] $.
Static disorder broadens $A(E,q)$. Its linewidth in $q$-space at fixed $E$ reveals information about the delocalized character of polariton modes at this energy, and whether $q$ is a good quantum number in the presence of disorder.\cite{ioffe1960non,Agranovich2003, Litinskaya2006,Ribeiro2022}

Complementary information is given by the molecular angle resolved DOS obtained from the vectors $|\beta(q)\rangle \equiv \frac{1}{\sqrt{N}} \sum_j e^{iq\cdot r_j}b^\dagger_j|0\rangle$. This quantity provides information on the matter component of  optically bright upper/lower polariton states at a given wavevector:
\begin{equation}\label{eq:beta_abs}
    \rho_\beta(E,q) \equiv \langle\beta(q)|\delta(H-E)|\beta(q)\rangle,
\end{equation}
which is evaluated stochastically analogously to Eq.  (\ref{eq:Awk}).  While in the presence of disorder $q$ is no longer a good quantum number, we clearly visualize (Fig. \ref{fig:ipr_3D}) the trade-off between molecular and photon contributions, and the energy broadening of the bright states in each subspace.

 Figure \ref{fig:agg_cav} shows the photonic density of states, 
 for a J-aggregate, and an identical lattice with no Coulombic intermolecular terms in the strong disorder regime $\sigma \sim J \sim \Omega_R $. As the number of photon modes is tiny compared to the number of molecular dipoles, there is essentially no change in the total DOS when the cavity is turned on. However, the $q\approx 0$ UP/LP states can be identified in the photon DOS when the J-aggregate is placed inside the microcavity.   

For all observables that include some form of internal aggregate structure there is exchange narrowing, i.e., interaction-induced narrowing of peaks and increase in the participation ratios.~\cite{Hestand2018} Figure \ref{fig:beta} shows that, in  J-aggregates, microcavity coupling induces substantially greater splitting between the LP/UP bands in the face of strong disorder, substantial exchange narrowing, and largely asymmetric line shapes skewing higher in energy. The additional narrowing (i.e., resistance to disorder), relates to the fact that the J-aggregate molecular DOS (Figure \ref{fig:agg_cav}(E)) extends higher in energy than an uncoupled system DOS, thus allowing higher energy photons to remain in resonance with the molecular system. The significant differences in line shape observed between the analyzed aggregates are entirely due to the delocalization of their respective dark exciton states as demonstrated in Figure \ref{fig:ipr_3D}.

 Figure \ref{fig:ipr_3D} shows the angle resolved molecular density of states and relative wavepacket uncertainty for a J-aggregate and noninteracting (uncoupled) emitters. We observe much greater wave character in the (weakly coupled) J-aggregate dark exciton modes at higher $q$, despite the influence of strong disorder. The enhanced wave character of the high $q$ weakly coupled modes is a byproduct of the strong intermolecular interactions in J-aggregates which also lead to the reduced photonic content in the J-aggregate LP band shown in Fig. \ref{fig:beta}(C) bottom. Additionally, as reported in the SI, the average participation ratio of J-aggregate molecular states is of order $10^4$, while an uncoupled polariton Hamiltonian leads to a maximum participation of 300. Lastly, the well-studied phenomena of exchange narrowing in molecular aggregates,\cite{SeeSI,Hestand2018,Anderson1953,Sumi1977} is also clearly visible in the $q\sim 0$ polariton transmission spectrum at the top of Fig. \ref{fig:beta}(C).

Overall, our results show that the long-range intermolecular interactions of organic aggregates lead to substantial effects in their (multimode) cavity-polariton spectra and dark state delocalization measures. This demonstrates the need to include accurate internal models for the electronic coupling in polaritonic systems made from crystals, polymers or aggregates. For example, the large Rabi splitting  (0.06 eV) obtained in the present 2D studies was the result of realistic $5-10$ D dipole and realistic molecular geometries. To attempt to produce such a Rabi splitting in a 1D lattice would require unrealistic dipoles on length scales where transition dipole coupling effects are no longer meaningful, leading to especially inadequate description of the dark modes. We expect this substantial delocalization of the molecular states to have important effects on photophysical properties and transport phenomena in these systems, leaving the door open for further studies that consider this effect.

The stochastically-evaluated observables are obtained here through an efficient molecular-coupling scheme that will apply to other,  more sophisticated model Hamiltonians, enabling future work to consider even more realistic system geometries and internal structure when studying energy and charge transfer in many-molecules polariton systems. 

\begin{acknowledgments}
    NCB acknowledges the NSF Graduate Research Fellowship Program under grant DGE-2034835. RFR acknowledges generous start up funds from Emory University. JRC was supported by NSF grant NSF CHE-2204263. DN was supported by NSF grant CHE-2245253.
\end{acknowledgments}

\bibliography{apssamp}
\end{document}